\documentclass[aps,prb,superscriptaddress,twocolumn]{revtex4}
\usepackage{graphicx}
\usepackage{amsmath}
\usepackage{bm}
\usepackage{xcolor}

\begin{document}

\title{Realization of valley-spin polarized current via parametric pump in monolayer $\rm \mathbf {MoS_{2}}$ }

\author{Kai-Tong Wang}
\affiliation{College of Physics and Optoelectronic Engineering, Shenzhen University, Shenzhen 518060, China}
\affiliation{School of Physics and Engineering, Henan University of Science and Technology, Luoyang 471023, China}

\author{Hui Wang}
\affiliation{School of Physics and Engineering, Henan University of Science and Technology, Luoyang 471023, China}

\author{Fuming Xu}
\email[]{xufuming@szu.edu.cn}
\affiliation{College of Physics and Optoelectronic Engineering, Shenzhen University, Shenzhen 518060, China}

\author{Yunjin Yu}
\affiliation{College of Physics and Optoelectronic Engineering, Shenzhen University, Shenzhen 518060, China}

\author{Yadong Wei}
\affiliation{College of Physics and Optoelectronic Engineering, Shenzhen University, Shenzhen 518060, China}

\begin{abstract}
Monolayer $\rm MoS_2$ is a typical valleytronic material with valley-spin locked valence bands. We numerically investigate the valley-spin polarized current in monolayer $\rm MoS_2$ via adiabatic electron pumping. By introducing an exchange field to break the energy degeneracy of monolayer $\rm MoS_2$, the top of its valence bands is valley-spin polarized and tunable by the exchange field. A device with spin-up polarized left lead, spin-down polarized right lead, and untuned central region is constructed through applying different exchange fields in the corresponding regions. Then, equal amount of pumped currents with opposite valley-spin polarization are simultaneously generated in the left and right leads when periodically varying two pumping potentials. Numerical results show that the phase difference between the pumping potentials can change the direction and hence polarization of the pumped currents. It is found that the pumped current exhibits resonant behavior in the valley-spin locked energy window, which depends strongly on the system size and is enhanced to resonant current peaks at certain system lengths. More importantly, the pumped current periodically oscillates as a function of the system length, which is closely related to the oscillation of transmission. The effects of other system parameters, such as the pumping amplitude and the static potential, are also thoroughly discussed.
\end{abstract}
\maketitle

\section{INTRODUCTION}

Valleytronics has attracted enormous attention on account of its potential for information processing \cite{Rycerz2007,Schaibley2016Valleytronics,Cheng2016The,GunlyckeGraphene,AngValley,Zhi2020Valley,ChenValley,Xiao2007,YuValley,ZhangInfluence,Niu2013Valley,
Zhang2014Generation,Paw2018Valley,Cheng2018Mani,Lei2020Valley,Gut2020Valley}. In many crystalline materials, there are two or more minima(maxima) at the conduction(valence) band in the momentum space, known as valleys. The degenerate but inequivalent valley states constitute new pseudospin degree of freedom for low energy carriers. Similar to spintronics, the essential of valleytronics is to generate and manipulate valley polarization to encode and store information. Various materials have been explored to realize valley polarization, including silicon\cite{TakashinaValley,CulcerValley}, bismuth\cite{Zhu2012Field}, diamond\cite{Nebel2013,IsbergGeneration}, carbon nanotube\cite{Fuming2016Transmission,AndrasDisorder}, etc. In particular, two-dimensional(2D) honeycomb lattice materials such as graphene or transition metal dichalcogenides (TMDs) provide a perfect platform to investigate valleytronics. Compared to graphene, TMDs labeled as $\rm MX_2$ $(\rm M=Mo, W, \rm X=S, Se, Te)$, also have two well-separated valleys in the Brillouin zone\cite{Xiao2007,XuSpin}. However, due to inversion symmetry breaking, TMDs are natural gapped semiconductors, which makes TMDs the promising candidates of valleytronic materials\cite{CaiMagnetic,Xiao2012,Junior2022First,SemenovBias,OliveiraValley,Tornatzky2018Resonance,Mak2014Valleytronics}.

As a typical TMDs material, monolayer $\rm MoS_2$ has a strong spin-orbit coupling(SOC) interaction\cite{Xiao2012,Zhu2011}, which leads to the locking between valley and spin at the top of its valance band. The valley-spin locking means that the valley and spin can be polarized together, and the lifetime of polarization can be enhanced due to the large spacing between $\rm K$ and $\rm K^{\prime}$ valleys. In the presence of an exchange field, TMDs exhibit interesting phenomena, such as the quantum anomalous Hall effect\cite{Yu2010,Zhou2017}, spin and valley Hall effects\cite{Da1112020}, and unconventional superconductivity\cite{Rahimi112017}. Besides, an exchange field can induce polarized valleys, which can be inverted by tuning the spin polarization. Through the ferromagnetic proximity effect\cite{Gan2013Two,Teng2020Mani} or magnetic doping\cite{Cheng2014,Zhou2017a}, the exchange field can be introduced into TMDs materials, which provides an effective way to manipulate its valley/spin degree of freedom. In experiments, the exchange field for valley splitting has been realized in Fe-doped\cite{Li2020Enha} or Co-doped monolayer $\rm MoS_2$.\cite{Zhou2020Syn} EuS as a ferromagnetic substrate can efficiently induce the magnetic exchange field in monolayer TMDs.\cite{Zhao2017Enh,Norden2019Gian}

Based on the valley optical selection rules, the optical pumping of valley polarization has been experimentally realized by circular polarized light in $\rm 2D$ TMDs\cite{Mak2012,Zeng2012,Mak2018Light}. Very recently, the spin-valley coupled dynamics at the $\rm MoS_2$-$\rm MoSe_2$ interface is experimentally studied using optical pumping\cite{Kumar2021Spin}; photoinduced valley-selective polarization in monolayer $\rm WS_2$ has been realized with circularly polarized light pumping.\cite{Guddala2021All} Besides, The line defects\cite{Pulkin2016Spin}, nonmagnetic disorders\cite{An2017}, and spatially varying potentials\cite{Gut2020Valley} were predicted to achieve the valley polarization in monolayer $\rm MoS_2$. In terms of applications, it is desirable to obtain pure valley polarized current by electrical methods. Accordingly, we propose that quantum parametric pump can drive valley and spin polarized currents in monolayer $\rm MoS_2$ through adiabatically varying two gate voltages. The parametric pump can produce dc current by periodically varying system parameters, which has been generalized to various $\rm 2D$ materials\cite{Talyanskii2001,Zhang2011,Wang2020Poss,Wang2022}. Specially, spin pump has been reported in several nanostructures\cite{Xing2004,Deng2015,Chen2015a}, where pure spin current and zero charge current are obtained.

\begin{figure}[tbp]
\centering
\includegraphics[width=\columnwidth]{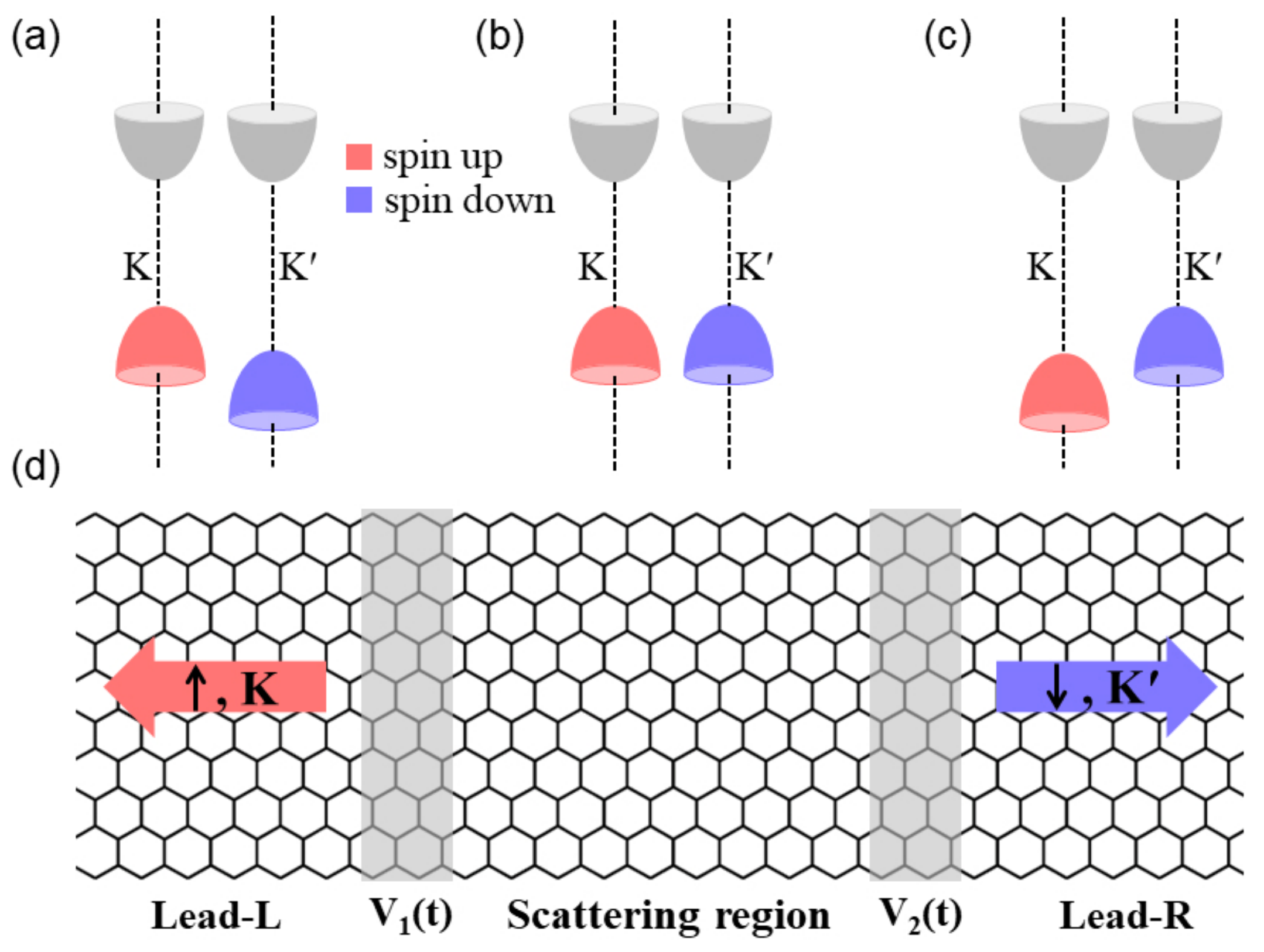}
\caption{Schematics of the band structures of monolayer $\rm MoS_2$ for (a) with exchange field $M$, (b) without exchange field, (c) with exchange field $-M$. The red and blue valance bands denote valley $K$ with spin up and $K^{\prime}$ with spin down. (d) The pump setup based on monolayer $\rm MoS_2$ consisting of left/right leads and the scattering region, whose band structures are correspondingly shown in (a) to (c), respectively. The $\rm MoS_2$ lattice is represented by the simple honeycomb lattice. The pumping potentials $V_1$ and $V_2$ are added in the scattering region, adjacent to the leads. As $V_1$ and $V_2$ periodically change, electric currents with opposite valley-spin polarizations are simultaneously pumped into the left and right leads, as shown by the block arrows.} \label{fig1}
\end{figure}

In this paper, we numerically study the generation and manipulation of valley-spin polarized currents via adiabatic pump in monolayer $\rm MoS_2$. The system setup is shown in Fig.\ref{fig1}. By magnetic doping, an exchange interaction is introduced in the left and right leads, which induces locked valley-spin polarization at the top of valence band as shown in Fig.1(a) and 1(c). When the pumping potentials periodically change, fully valley-spin polarized dc currents are driven into the leads. At one moment, the current with $\rm K$ valley and spin up is pumped into the left lead while the current with opposite valley-spin polarization flows into the right lead. The polarized current exhibits resonant behavior in the valley-spin locked energy window, which mainly depends on the system size. With the increasing of the system length, the pumped currents show periodic oscillation behavior and robust resonant current peaks can be observed. We also investigate the influence of other system parameters, including the phase difference, the Rashba SOC strength, the static potential, and the Fermi energy. It is found that the phase difference and static potentials can invert the direction and hence polarization of the pumped current.

The paper is organized as follows. In Sec. \uppercase\expandafter{\romannumeral2}, we introduce the Hamiltonian of monolayer $\rm MoS_{2}$ and the formalism of adiabatic parametric pumping. In Sec. \uppercase\expandafter{\romannumeral3}, numerical results and relevant discussions are presented. Finally, a brief summary is given in Sec. \uppercase\expandafter{\romannumeral4}.

\section{Model and formalism}

In monolayer $\rm MoS_{2}$, the low-energy spectrum at $\rm K$ and $\rm K^{\prime}$ valleys consists of three $d$ orbitals of $\rm Mo$, i.e., $d_{z^2}, d_{x^2-y^2}, d_{xy}$. The relations between these orbitals and basis wave functions satisfy: $|\varphi_{c}\rangle=|d_{z^2}\rangle, |\varphi_{\upsilon}^{\lambda}\rangle=  (|d_{x^2-y^2}\rangle + i\lambda |d_{xy}\rangle)/\sqrt{2}$, where the subscript $c/\upsilon$ denotes the conduction/valence band and $\lambda=\pm 1$ corresponds to different valleys $\rm K$ and $\rm K^{\prime}$. Based on above low-lying states, the effective Hamiltonian of monolayer $\rm MoS_{2}$ has the following form\cite{Xiao2012,Ochoa2013}
\begin{equation}\label{}
H_{0}(k) = at(\lambda k_{x}\sigma_{x}+k_{y}\sigma_{y}) + \Delta \sigma_{z} - t_{SO}\lambda \frac {\sigma_{z}-1}{2} \tau_{z},
\end{equation}
where $a$ and $t$ are the lattice constant and hopping strength, respectively. $\sigma_{x,y,z}$ and $\tau_z$ represent the Pauli matrices of basis functions($|\varphi_{c}\rangle$ and $|\varphi_{\upsilon}\rangle$) and spin($\uparrow$ and $\downarrow$). $\Delta$ is the mass term and the last term is the intrinsic SOC with strength $t_{SO}$.

We employ the tight-binding model of $\rm MoS_{2}$, which treats monolayer $\rm MoS_{2}$ as a simplified honeycomb lattice. The lattice includes A and B sublattices, corresponding to the $d_{z^2}$ orbit and $d_{x^2-y^2} + i\lambda d_{xy}$ orbits of Mo, respectively. In the tight-binding approximation, the Hamiltonian can be expressed as\cite{Klinovaja2013,Kleftogiannis2015}
\begin{equation}\label{}
H_{0} = \sum_{i}\epsilon_{i}c_{i\alpha}^{\dag}c_{i\alpha} + t\sum_{<i,j>}c_{i\alpha}^{\dag}c_{i\alpha} + H_{SO},
\end{equation}
with
\begin{equation}\label{}
H_{SO} = \frac{2it_{SO}}{3\sqrt{3}} \sum_{\ll i,j \gg,\alpha,\alpha^{\prime}} \upsilon_{ij}c_{i\alpha}^{\dag} \tau_{z,\alpha\alpha^{\prime}} c_{j\alpha^{\prime}},
\end{equation}
where $c_{i\alpha}^{\dag}(c_{i\alpha})$ is the creation(annihilation) operator at site $i$ with spin $\alpha=\pm 1$, $\epsilon_{i}$ is the on-site energy. $H_{SO}$ denotes the intrinsic SOC term and the summation over the second nearest-neighbor sites only involves B sublattice. Besides, $\upsilon_{ij} =  +1(-1)$ if an electron moves from site $j$ to site $i$ with taking a left(right) turn\cite{Kane2005}.

Based on this model, we consider a monolayer $\rm MoS_{2}$ setup, which contains three parts: the central scattering region, the left and right leads, as shown in Fig.1(d). By magnetic doping, different valley-spin polarizations can be induced in the left and right leads due to the exchange field\cite{Cheng2014,Bai2018}. The schematic band structures with or without the exchange field are depicted in Fig.1(a)-(c). In the presence of Rashba spin-orbit coupling (RSOC), the Hamiltonians for the scattering region with pumping potentials and the leads can be written as
\begin{equation}
H_{C} = H_{0} + \frac{3it_R}{4} \sum_{<i,j>,\alpha,\alpha^{\prime}} (\mathbf{\tau}_{\alpha \alpha^{\prime}}\times \mathbf{d}_{ij})_{z} c_{i\alpha}^{\dag} c_{j\alpha^{\prime}} + V(x,y,t), \  \\
\end{equation}
\begin{equation}
H_{L/R} = H_{0} \pm M\sum_{i,\alpha,\alpha^{\prime}} \tau_{z,\alpha\alpha^{\prime}} c_{i\alpha}^{\dag}c_{i\alpha^{\prime}}, \\
\end{equation}
where $M$ and $t_{R}$ denote the strengths of the exchange field and RSOC, respectively. $\tau_{\alpha \alpha^{\prime}}=(\tau_x,\tau_y,\tau_z)$ is the Pauli matrix for spin, and $\mathbf{d}_{ij}$ is the lattice vector connecting sites $i$ and $j$. The potential term $V(x,y,t) = V_{s}(x,y)+V_{t}(x,y,t)$, where $V_{s}=V_{0}\sum_{i} \Pi_{i}(x,y)$ corresponds to the static potential defining the shape of the pumping region. $V_{t}=V_{p}\sum_i \Pi_{i}(x,y)cos(\omega t +\varphi_{i})$ is the periodic pumping potential. $V_{0}$ and $V_{p}$ are the amplitudes of $V_{s}$ and $V_{t}$. $i=1,2$ are the indices of the potential and $\Pi_i$ represents the potential profile, which is highlighted in green in Fig.1(d). $\varphi_{i}$ is the initial phase of the pumping potential.

To evaluate the adiabatic valley-spin pump, we need to calculate the average current flowing into lead $\beta$. Consider a slowly varying time-dependent pumping potential $V_{t,i}$, the average current in one period is expressed as\cite{Brouwer1998}
\begin{equation}\label{}
I_{\beta} = \frac {q\omega}{2\pi}\int_{0}^{T}dt[\frac {dN_{\beta}}{dV_{t,1}} \frac {dV_{t,1}}{dt}+\frac {dN_{\beta}}{dV_{t,2}} \frac {dV_{t,2}}{dt}],
\end{equation}
where the period of $V_{t,i}$ is $T= 2\pi / \omega$ with frequency $\omega$ and $\beta=L/R$ labels the lead. The emissivity $\frac{dN_{\beta}}{dV_{i}}$ is defined in terms of the scattering matrix $S_{\beta\beta^{\prime}}$ as\cite{Buttiker1994,Wei2000}
\begin{equation}\label{}
\frac{dN_{\beta}}{dV_{i}} = \int \frac{dE}{2\pi}(-\partial_{E}f)\sum_{\beta^{\prime}}Im \frac{\partial S_{\beta\beta^{\prime}}}{\partial V_{i}} S_{\beta\beta^{\prime}}^{\ast},
\end{equation}
with $f$ the Fermi distribution function. Under the adiabatic condition, the pumped current is independent of the pumping frequency $\omega$, hence we set $\omega=1$ in the calculation.

In the language of nonequilibrium Green's functions, the pumped current is expressed as\cite{Wang2003,Xu2011,Yu2016}
\begin{equation}
I_{\beta} = -\frac{q}{2\pi}\int_{0}^{2\pi}dt\int dE (\partial_{E}f) Tr[\Gamma_{\beta}G^{r}\frac{dV_{t}}{dt}G^{a}].
\end{equation}
Here $G^{r}$/$G^{a}$ is the retarded/advanced Green's function of the central scattering region, which is defined as $G^{r}=G^{a,\dag}=[E-H_{C}-\sum_{\beta}\Sigma_{\beta}^{r}]^{-1}$. $H_{C}$ is the corresponding Hamiltonian. $\Sigma_{\beta}^{r}$ is the retarded self-energy of lead $\beta$, which can be calculated by surface Green's function\cite{Lee1981,Lee1981a}. $\Gamma_{\beta}=i(\Sigma_{\beta}^{r}-\Sigma_{\beta}^{a})$ denotes the linewidth function.

As shown by block arrows in Fig.1(d), at one moment of the pumping period, polarized current with $K$ valley and spin up (pink arrow) is driven into the left lead, and equal amount current with $K'$ valley and spin down (blue arrow) flows in the right lead. Detailed numerical results are shown in the following section. We use the short term, the pumped current, to stand for the valley-spin polarized currents in the leads.

\section{Results and discussion}

In the calculations, the on-site energy is $\epsilon_{i}=\pm 0.83~\rm eV$ for A and B sublattices. Other parameters\cite{Xiao2012,Klinovaja2013} are set as $t=1.27~\rm eV$, $t_{SO}=0.038~\rm eV$, and the lattice constant a=0.32~nm. Without loss of generality, we set the exchange field strength $M=0.06~\rm eV$, and $\rm eV$ is taken as the energy unit throughout the calculation. The periodic boundary condition (PBC) is considered for monolayer $\rm MoS_{2}$, and thus the edge effect is removed. To realize PBC, the upper and lower edges of a zigzag $\rm MoS_{2}$ ribbon shown in Fig.1(d) are connected with appropriate hopping interactions, which is also called the cylinder boundary.

\begin{figure}[tbp]
\centering
\includegraphics[width=0.9\columnwidth]{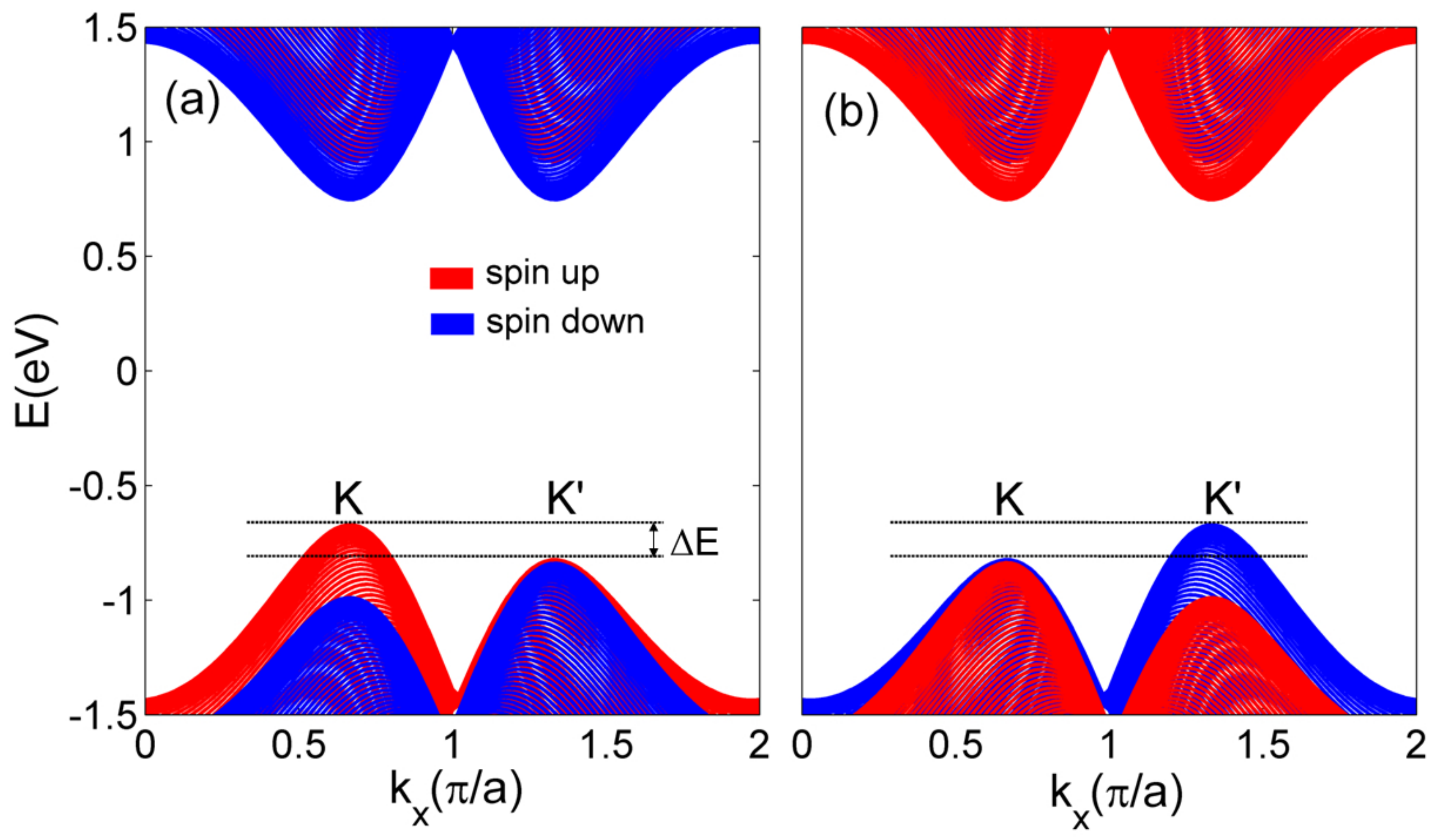}
\caption{The dispersion relation of monolayer $\rm MoS_2$ ribbon with an exchange field: (a) $M=0.06$, (b) $M=-0.06$. Both valley and spin are polarized together, where $\rm\Delta E$ is defined as the valley-spin locked energy window.}\label{fig2}
\end{figure}

\subsection{Valley-spin polarized current}

The dispersion of monolayer $\rm MoS_{2}$ with an exchange field is plotted in Fig.2. From Fig.2(a), it is clear that the valley $\rm K$ with spin up is polarized at the valence band top. However, as the exchange field changes, the polarization of both valley and spin is inverted as shown in Fig.2(b). In the valley-spin locked window $\Delta E$, perfect polarization can be realized. To investigate the valley-spin polarized current, we consider only the $\Delta E$ energy range. In Fig.3, We study the dependence of the pumped current on the phase difference $\varphi_{12}$ between $V_{1}$ and $V_2$. Due to the inverse valley-spin polarization of the left and right leads, the holes are forbidden to propagate through the scattering region, so there is no pumped current generated in this setup at $t_{R}=0$. Introducing RSOC in the scattering region, it is found that the pumped current arises and flows into left or right lead. We attribute such a dc current to the spin flip process induced by the RSOC. Importantly, nonzero valley-spin polarized currents are pumped into different leads, which depends on the exchange field in leads. Our results show that $I_{L,R}$ is an odd function about $\varphi_{12}$, i.e., $I(\varphi_{12})=-I(-\varphi_{12})$. The maximum of the pumped current appears at $\varphi_{12}=\pm \pi/2$.

\begin{figure}[tbp]
\centering
\includegraphics[width=0.9\columnwidth]{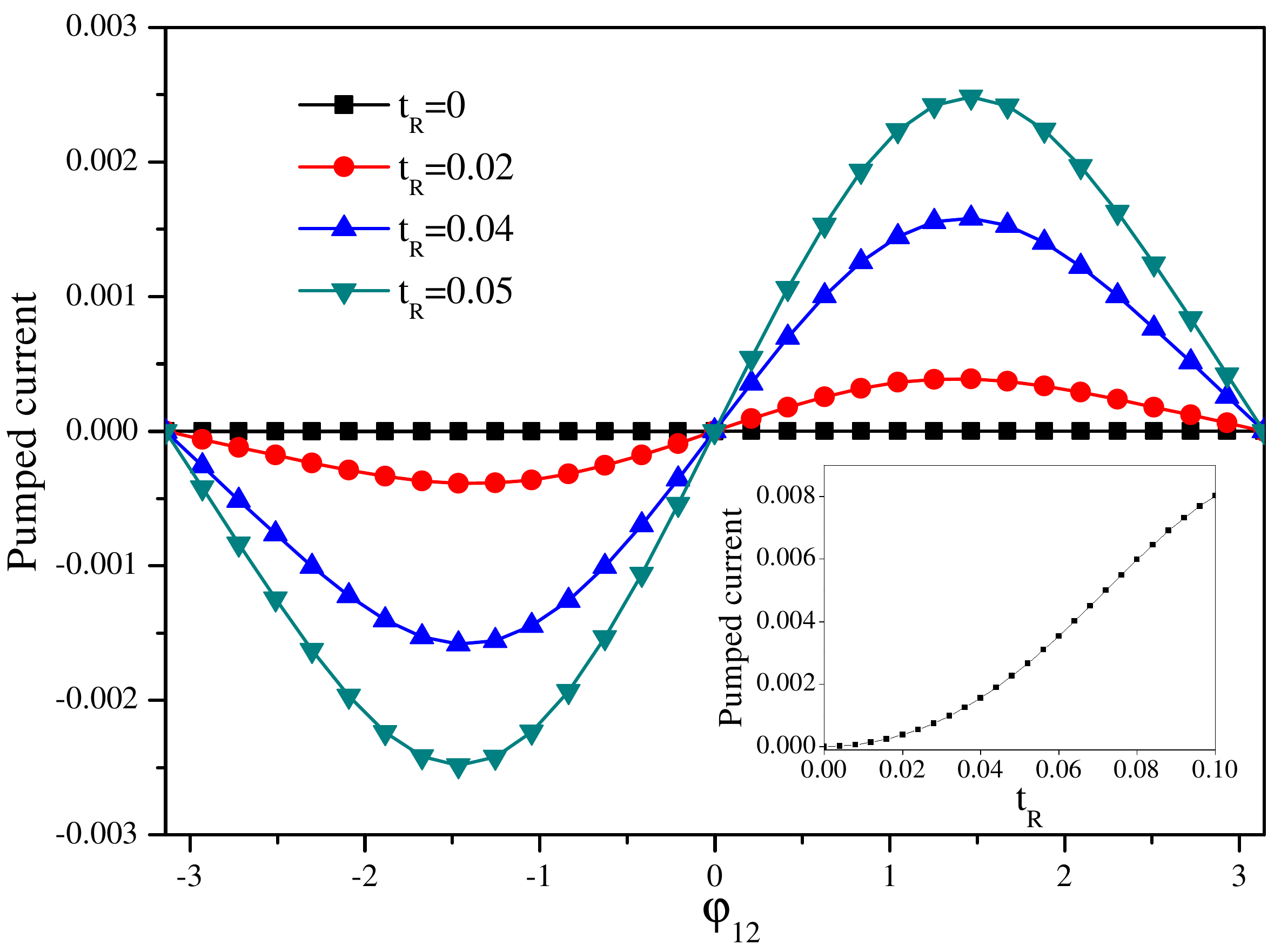}
\caption{The pumped current $I_{L}$ as a function of the phase difference $\varphi_{12}$ between $V_1$ and $V_2$. Inset: the pumped current versus RSOC strength $t_R$ at $\varphi_{12}={\pi}/{2}$. Other parameters: $E_f=-0.74$, $L=10a$, $V_0=0.1$, $V_p=0.05$. }\label{fig3}
\end{figure}

From Fig.3, when the phase difference $\varphi_{12}$ shifts from $-\pi$ to $0$, the valley-polarized holes with spin up will be pumped out of the scattering region and flow into the left lead. On the contrary, the opposite valley-polarized holes with spin down will spread into the right lead when $\varphi_{12}$ shifts from $0$ to $\pi$. It means that the direction of the pumped current can be tuned by the phase difference $\varphi_{12}$, then different valley-spin polarized currents are pumped into different leads. We calculate the pumped current as a function of the phase difference $\varphi_{12}$ for different RSOC strengths $t_{R}$. Consequently, with the increasing of $t_{R}$, the pumped current increases. $I_{L}$ versus $t_{R}$ at $\varphi_{12}=\frac{\pi}{2}$ is plotted in the inset. The result is understandable: since the spin-flip efficiency increases when $t_{R}$ is increased, the spin-up carriers from one lead can be more easily flipped as spin-down carriers and flow into the other lead, which is required by the conservation of charge.

\begin{figure}[tbp]
\centering
\includegraphics[width=0.9\columnwidth]{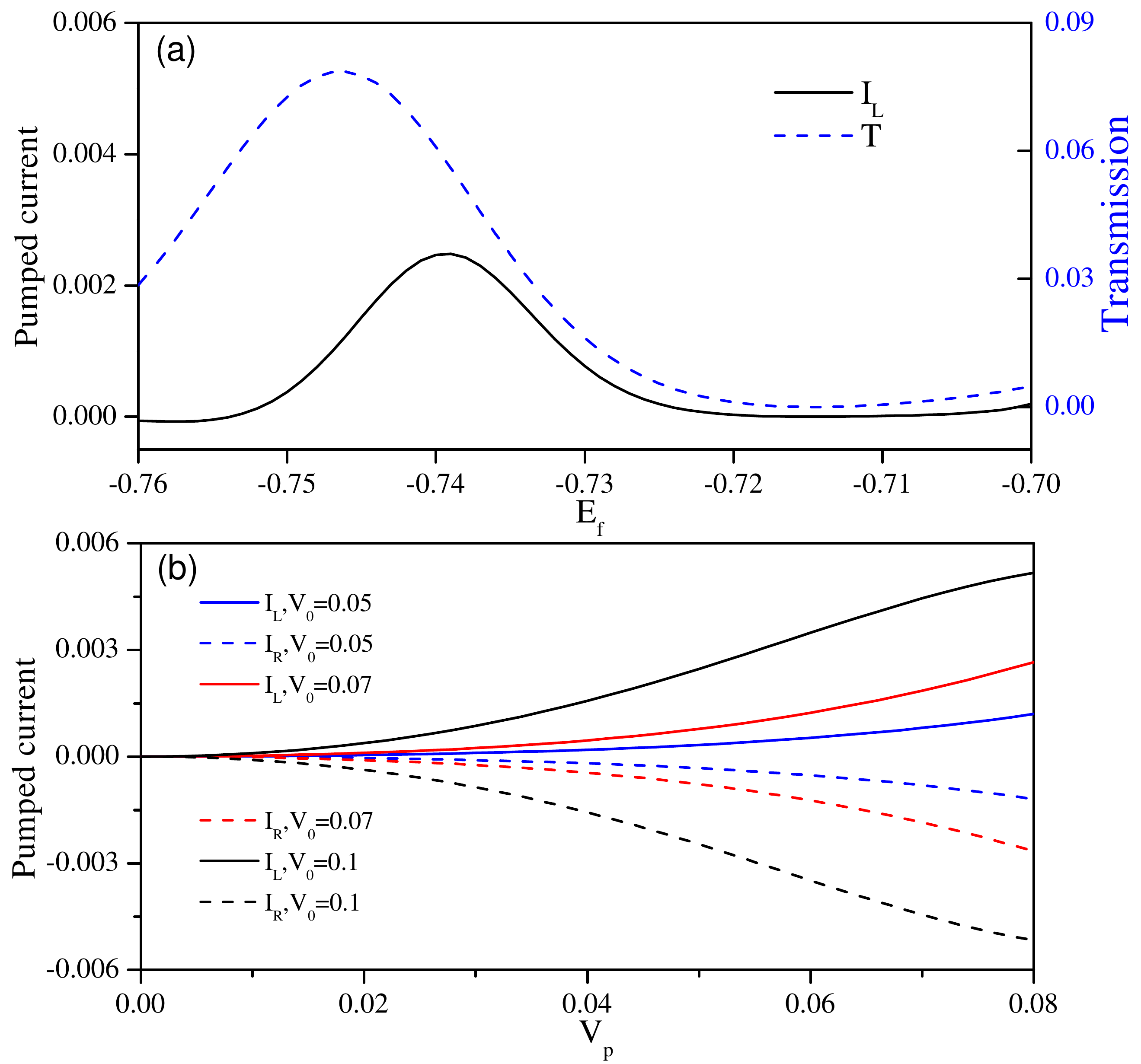}
\caption{(a) The pumped current and transmission versus Fermi energy at $V_0=0.1$ and $V_p=0.05$. (b) The polarized current $I_L$ and $I_R$ versus the pumping potential $V_p$ for different static potentials $V_0$ at $E_f=-0.74$. Other parameters: $\varphi_{12}={\pi}/{2}$, $L=10a$, $t_{R}=0.05$. }\label{fig4}
\end{figure}

In Fig.4(a), the pumped current and the transmission versus Fermi energy $E_f$ are plotted at $\varphi_{12}=\frac{\pi}{2}$. Obviously, a broad transmission peak arises in the locked window $\Delta E$, and the pumped current exhibits similar behavior as the transmission. This result indicates that the transport of polarized holes is dominated by quantum resonance, which originates from quantum interference effect. In fact, the resonance assisted transport is a common property of electron pump\cite{Wei2000}. Besides, an impassable interval for the pumped current is generated as the Fermi energy is away from the resonant peak as shown in Fig.4(a). The variation of polarized current is well correlated to the transmission. The pumped current $I_{L,R}$ versus the pumping potential for different static potentials is plotted in Fig.4(b). Due to the particle conservation, the current flowing into the scattering region must satisfy $I_{L}=-I_{R}$, which is confirmed through the symmetric curves of $I_{L}$ and $I_{R}$. Furthermore, the results show that the valley-spin polarized current linearly increases with the increasing of pumping potential. For a relatively small $V_{p}$, the static potential $V_{0}$ can enhance the magnitude of pumped currents.

\begin{figure}[tbp]
\centering
\includegraphics[width=\columnwidth]{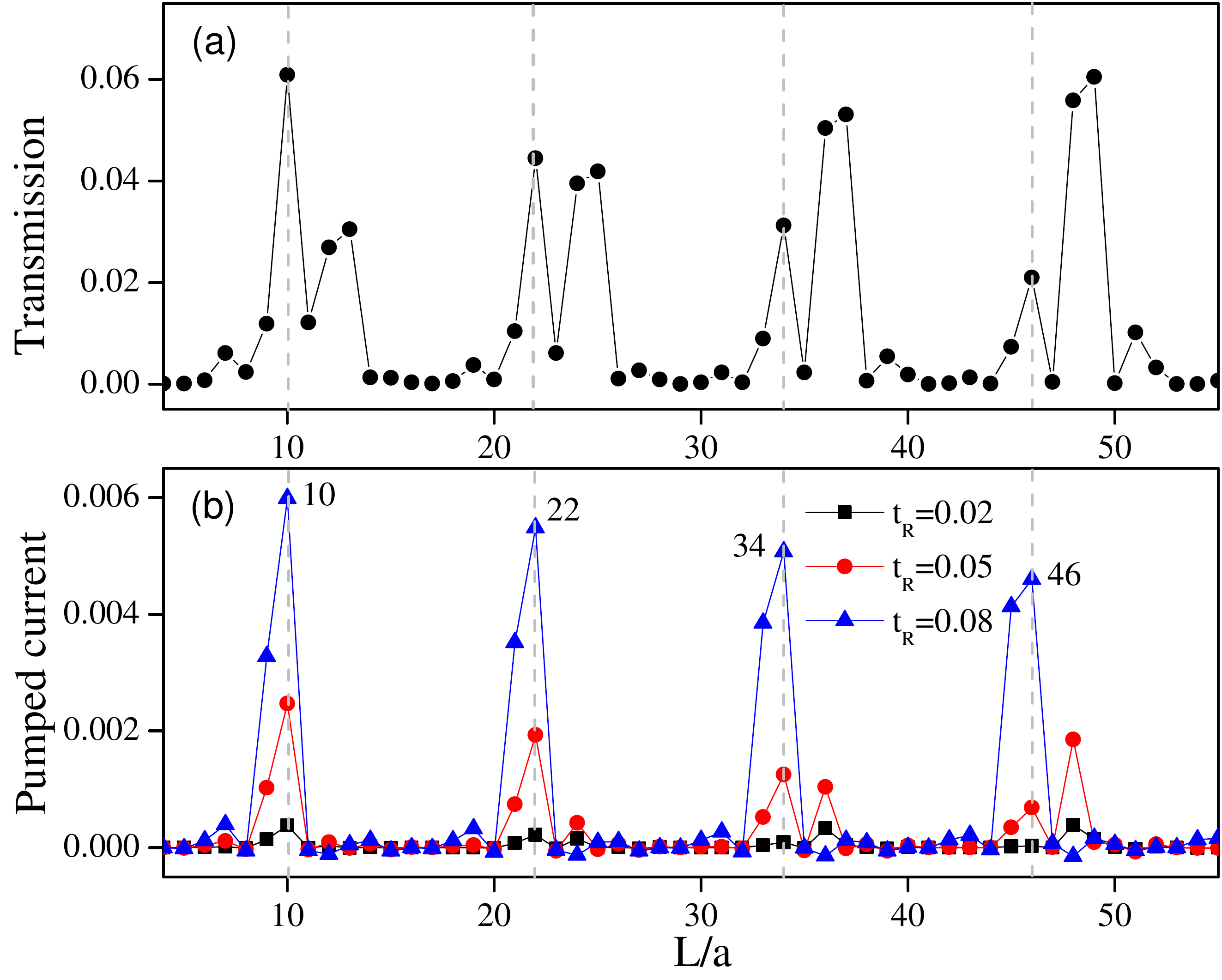}
\caption{(a) The transmission versus the length $\rm L$ of scattering region for $t_R=0.05$. (b) The pumped current $I_L$ as a function of system length $L$ for different RSOC strengths at $\varphi_{12}={\pi}/{2}$. The numbers label the lengths of the scattering region where current peaks emerge. Other parameters are the same as Fig.4.} \label{fig5}
\end{figure}

\subsection{Size effect on the pumped current}

In the following, we study the influence of the system size on the pumped current. Transmission and the pumped current versus the length of the scattering region are plotted in Fig.5. It is interesting that some robust peaks of the pumped current appear at certain lengths, but the currents for other lengths are almost zero. Moreover, these peaks show periodic oscillation behavior with the period length $12a$. In Fig.5(a), we plot the transmission as a function of the length $L$ for $t_R=0.05$. The transmission exhibits a similar behavior as the pumped current, where the transmission peaks also appear at certain system lengths. It is clear that these transmission peaks correspond exactly to the pumped current peaks. It is reasonable that the periodic behavior of the pumped current originates from the spin precession\cite{Yang2008Spin,Culcer2006Spin,Chen2021Time} induced by Rashba SOC. When the current carriers travel through the central scattering region with RSOC interaction, carrier spin keeps precessing and spin flip occurs, which results in the periodic oscillating behavior of the pumped current. Our calculation further demonstrates that the width of scattering region has almost no influence on the periodic behavior of polarized current as long as Fermi energy lies in the valley-spin locked energy window. To evaluate the influence of RSOC, We show in Fig.5(b) the pumped currents for different RSOC strengths. It is clear that the RSOC strength has less influence on the resonant period, which is $L=12a$. However, with the increasing of $t_R$, the resonant current peaks become significant. The peak current value grows larger as $t_R$ increases, which is consistent with the results shown in Fig.3.

Notice that this periodic oscillation behavior of the pumped current is different from the even-odd conductance oscillation of carbon-atom chains.\cite{Lang1998,Lang2000} When two metallic electrodes are attached to a carbon atomic chain, the electronic structure of the carbon chain is modified, which leads to the difference in the density of states between odd- and even-number carbon chains.\cite{Lang1998,Lang2000} This leads to the even-odd conductance oscillation driven by dc bias. However, for the periodic oscillation of the valley-spin polarized currents, it is due to the spin-flipping process induced by Rashba SOC and driven by periodic pumping potentials.

\begin{figure}[tbp]
\centering
\includegraphics[width=0.9\columnwidth]{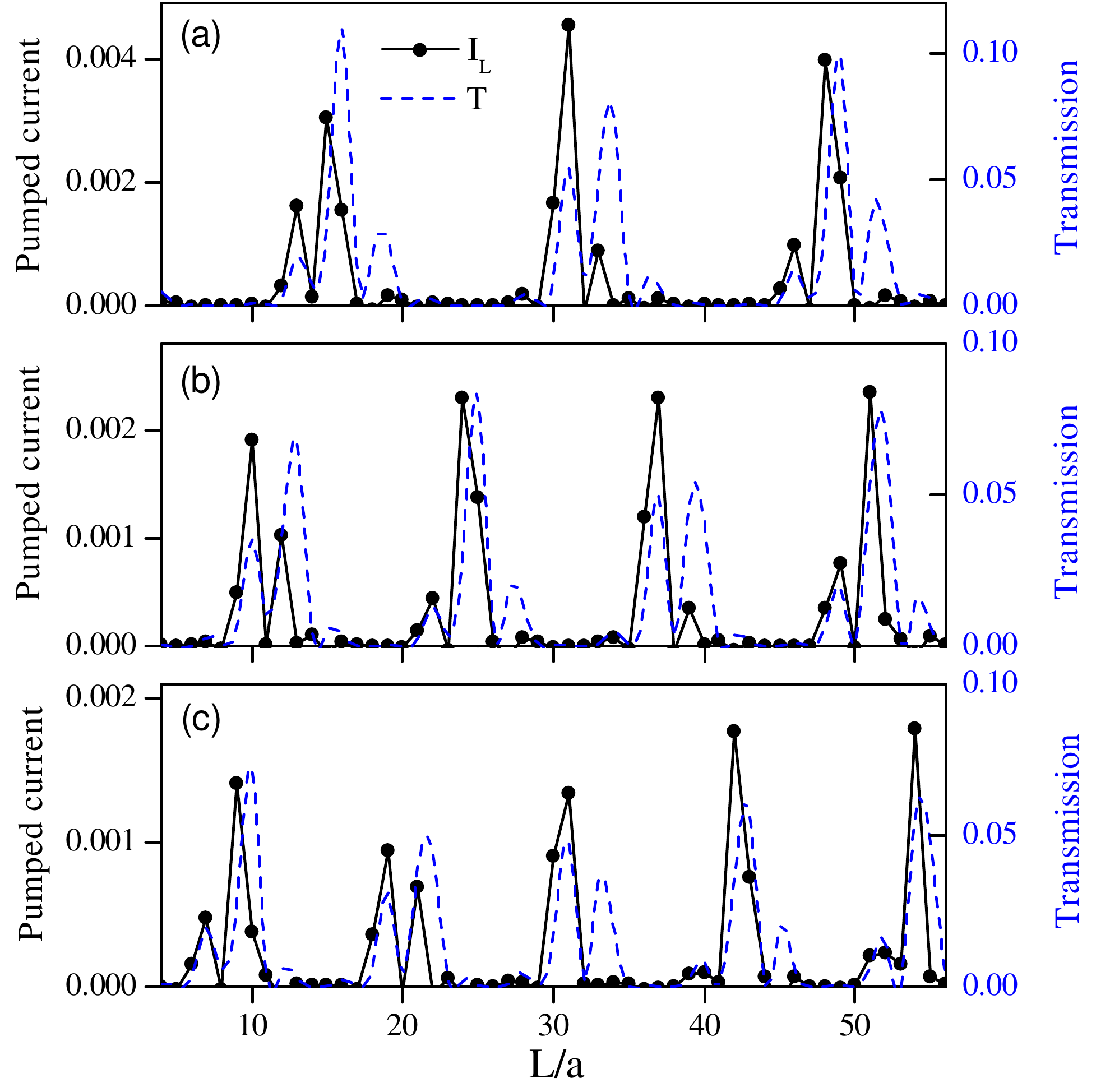}
\caption{The pumped current and transmission versus the length $L$ of the scattering region for different Fermi energies. (a): $E_f=-0.72$, (b): $E_f=-0.735$, (c): $E_f=-0.75$. Other parameters are the same as Fig.4.} \label{fig6}
\end{figure}

In Fig.6, we plot the pumped current and transmission versus the system length for different Fermi energies. Apparently, the periodic oscillation behavior of the pumped current persists as the Fermi energy changes. The result shows that, with the increasing of $E_{f}$, the number of peaks increases while the corresponding periodic length decreases. Besides, the magnitude of current peaks will decrease as the Fermi energy increases. By calculating the transmission, it can be seen that the behavior of the pumped current is still consistent with the transmission, which means the periodic oscillation is a universal phenomenon in the setup.

\begin{figure}[tbp]
\centering
\includegraphics[width=0.9\columnwidth]{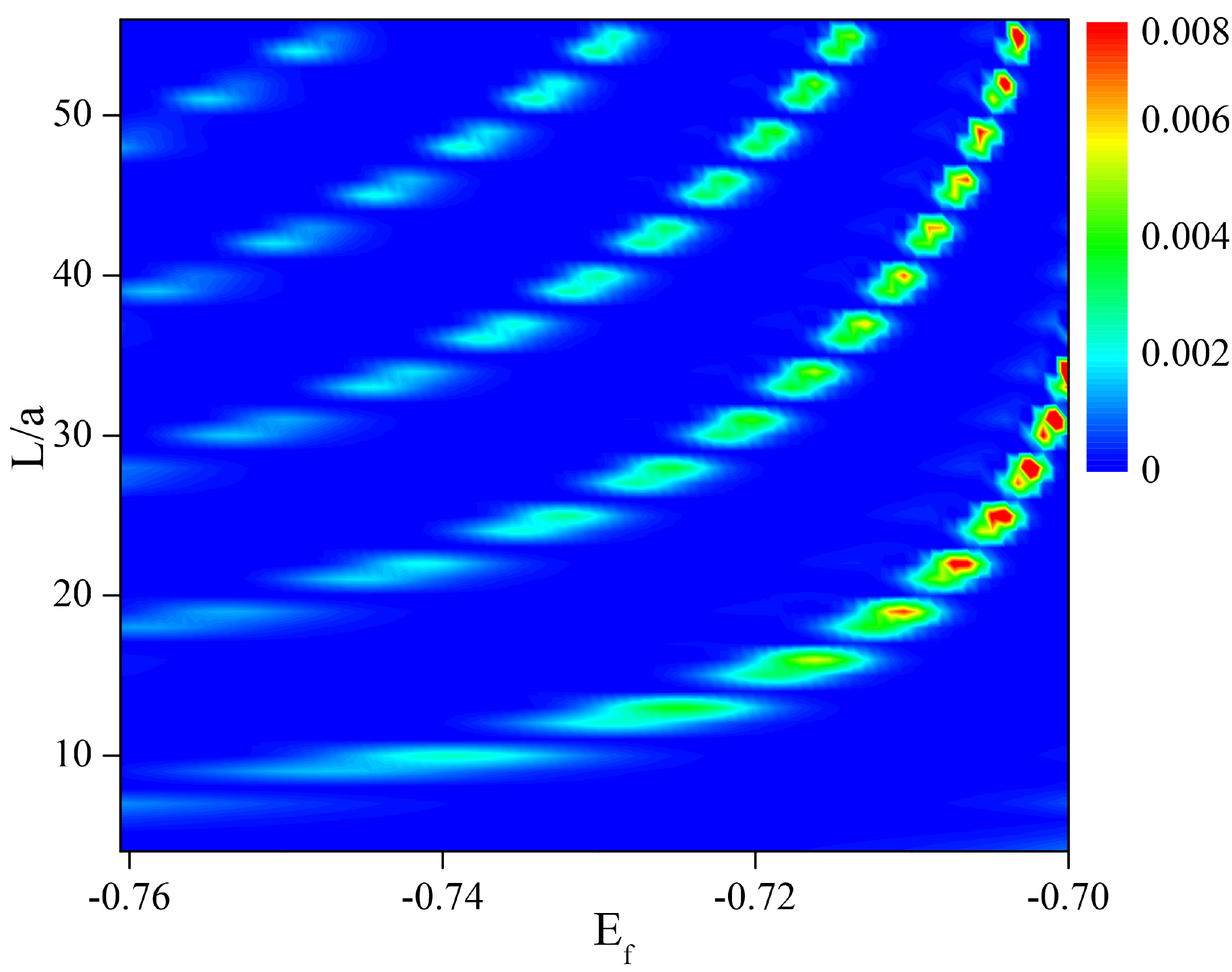}
\caption{The pumped current with respect to both the system length $L$ and the Fermi energy $E_f$.}\label{fig7}
\end{figure}

To exhibit an overall view of the pumped current, in Fig.7, we provide a two-dimensional diagram of the valley-spin polarized current as a function of both the system length $L$ and the Fermi energy $E_f$. By varying $L$ and $E_f$, we find that there are five curves with discrete extrema of currents. The periodic dependence of the pumped current on the system length $L$ is clearly shown. When $E_f$ approaches the top of valley, the largest pumped current appears and becomes more sharp as shown by the orange regions. Moreover, it is further confirmed that, multiple resonant peaks of the pumped current arise under an appropriate system length $L$. Our numerical results reveal that it is possible to design a high-efficiency device setup for generating valley-spin polarized currents.

\subsection{Influence of the static potential}

\begin{figure}[tbp]
\includegraphics[width=\columnwidth]{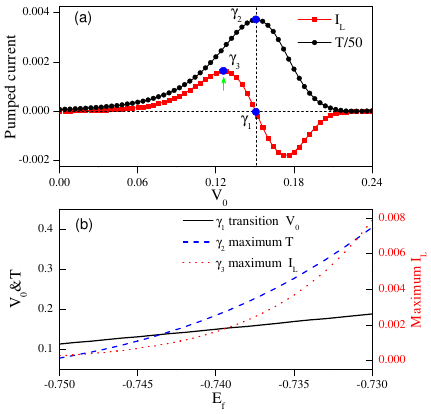}
\caption{(a) The pumped current as well as the transmission coefficient $T$ as a function of the static potential $V_0$. A factor of 1/50 is multiplied to T for better illustration. $\gamma_{1}$, $\gamma_{2}$, $\gamma_{3}$ label the critical points. (b) The maximum of $I_{L}$, transition point of $V_0$ and corresponding maximum of $T$ versus the Fermi energy. Other parameters: $V_p=0.03$, $\varphi_{12}= {\pi}/{2}$, $L=10a$. }\label{fig8}
\end{figure}

In this section, we focus on the dependence of $I_L$ on the static potential $V_{0}$. For this purpose, the system length and the RSOC strength are fixed at $L=10a$ and $t_R=0.05$. In Fig.8(a),  we plot the pumped current as well as transmission coefficient versus the static potential $V_{0}$. With the increasing of the static potential, a current peak first emerges at $V_0=0.126$ labeled by the blue point $\gamma_{3}$. As $V_0$ scans the critical point $\gamma_{1}$ at $V_0=0.151$, the direction of $I_L$ is reversed. Continuing to increase $V_0$, we can see a negative current peak. The result shows that the static potential can also change the direction and hence the polarization of the pumped current. Besides, in the vicinity of the critical point $\gamma_{1}$, a transmission peak is clear, which suggests the influence of the static potential also results from quantum resonance.

In Fig.8(b), The critical point of $V_0$ and the maxima of both $T$ and $I_L$ versus Fermi energy $E_f$ are plotted. It is found that the critical value of $V_0$ increases linearly with the increasing of $E_f$, which indicates the resonant energy level depends on the static potential. Besides, the curves of the peak values for transmission and pumped current grow with the Fermi energy. The variation of the pumped current with the Fermi energy is consistent with the results in Fig.6.

In this work, the valley-spin polarized current is generated by electric pumping in adiabatic regime, which requires two independently varying system parameters. We emphasize that similar mechanism can also be achieved with optical pumping, which is in the non-adiabatic regime due to the high frequency of light wave. In this case, the light frequency can serve as a pumping parameter. Therefore, non-adiabatic parametric pumping using electric or optical ways will certainly bring in more physics.

\section{Conclusions}
In conclusion, we study the valley-spin polarized current in monolayer $\rm MoS_2$ ribbon via parametric electron pump. In the proposed setup, different valley-spin polarized currents can be controlled to flow into different leads in the valley-spin locked energy window. The phase difference between the pumping potentials can change the direction and hence polarization of the pumped current, where quantum resonance dominates the transport process. Furthermore, the size effect on the valley-spin polarized current is numerically investigated. As the length of the scattering region changes, resonant peaks of the pumped current arise and show periodic oscillation due to the spin precession. With the increasing of Fermi energy, the number of peaks decreases while the peak height increases within a fixed length range. The dependence of the resonant pumped current on the scattering region length and the Fermi energy is numerically revealed in a two-dimensional diagram. It is also found that the direction of valley-spin polarized currents can be inverted by the static pumping potential.
As a potential valleytronic device, the pump setup proposed in this work can serve as a valley-spin polarization source, which can simultaneously generate opposite polarization in one device. The polarized signals can be efficiently manipulated by many system parameters, and significant resonant enhancement has been demonstrated.

$${\bf ACKNOWLEDGMENTS}$$

This work was supported by the National Natural Science Foundation of China (Grant Nos. 12034014 and 61674052) and the Natural Science Foundation of Shenzhen (Grant Nos. 20200812092737002, JCYJ20190808115415679, and JCYJ20190808152801642). Hui Wang also acknowledges supports from the Outstanding Youth Foundation of Henan Scientific Committee (212300410041) and the Key Scientific and Technological Projects in Henan Province (212102210223).


\end{document}